\documentclass[preprint]{aastex}
\usepackage{natbib}

\begin{document}

\title{On the Power Spectrum Density of Gamma Ray Bursts}
\author{Motoko Suzuki and Masahiro Morikawa \and Izumi Joichi}

\affil{
Department of Physics, Ochanomizu University, Tokyo 112--8610, Japan }

\affil{
School of Science and Engineering,
Teikyo University,
Toyosatodai 1--1, Utsunomiya 320--8551, 
Japan}

\begin{abstract}

Gamma ray bursts (GRBs) are known to have short-time variability
and power-law behavior with the index $-1.67$ in the power
spectrum density.
Reanalyzing the expanded data, we have found 
a) the power-law comes from
the global profile of the burst and not from the self-similar
shots nor rapid fluctuations in the luminosity profile. 
b) The power indices vary from burst to burst and the
value $-1.67$ is given simply as the mean value of the
distribution; there is no systematic 
correlation among GRBs to yield the power law. 
\end{abstract}

\section{
INTRODUCTION}

The origin and the fundamental mechanism of the gamma-ray
bursts (GRBs) have not been revealed despite the fact that
more than thirty years have passed since the first 
discovery of them.  In the numerous luminosity profiles of
GRBs, we notice that most 
GRBs have very rapid milli-second time variability in their intensity
\citep{Walker98} though there is
vast variety in luminosity profile itself for each
GRB. We believe that this characteristic variability must have
important information to reveal the GRB mechanism.
Therefore, in this letter, we would like to concentrate on
the analysis of this variability.  

In studying time sequence of objects, the Fourier transform
technique is useful and often yields indispensable information on the
scaling properties and characteristic time scale. For example,
\citet{Beloborodov98} used
this method for 214 light curves of
long GRBs ($T_{90}>20\sec $) and reported that the averaged
power spectrum density (PSD) of GRBs shows the power-law
with the index $1.67\pm 0.02$ over two decades in frequency range.
This value is very closed to $-5/3$ which
suggests the Kolmogorov spectrum of velocity fluctuations in
turbulent medium.

On the other hand, the power-law in PSD does not necessarily specify
the whole mechanism as we know in various examples.
Actually the power-law in PSD can be derived by many reasons; some
specific burst profile in luminosity, Levy-type random noise
in the background, and the superposition of similar shots and so on.
For example, the power-law in PSD of X-ray emission
from the Active Galactic Nuclei (AGN) is considered to be originated from
the superposition of many similar
shots \citep{Mineshige01}. \citet{Chang00}
tried to reproduce the power-law index, -5/3, and the individual
PSD distribution of GRBs by the superposition of many decaying pulse shots.  

Here in this paper, we would like to determine the origin
of the power-law in PSD of GRBs by analyzing the detail of PSD
for each burst data.

First, we demonstrate that the individual PSD of GRBs does
not exactly have the
power-law index $-1.67$ but has wide variation which is
well approximated by the Gaussian distribution.
Second, we argue that power-law index of PSD is determined by the 
profiles of an individual shot in the light curve and not by the 
superposition of many similar shots nor by
Levy-type random noise. Then we discuss the general nature 
of PSD of superposed shots and apply this method to the actual data 
of GRB light curves. Finally we observe how the averaged PSD shows 
clear power-law index $-1.67$.

\section{POWER LAW IN PSD OF GRB}

In our analysis, we use the data detected by the Burst and Transient
Source Experiment (BATSE) on Compton Gamma Ray Observatory (CGRO) 
with 64ms resolution
\footnote{
\url{ftp://cossc.gsfc.nasa.gov/pub/data/batse/ascii\_data/64ms/}}.
We use the light curves in the energy band $50< h\nu < 100$ keV,
excluding unclear high and low energy bands, 
and subtract the background by the linear fitting. Then
for excluding the noise contamination, we select the data with
peak count rate larger than
250 counts per 64ms bin; finally 297 data set remain.

In order to see the generality of the power index $-5/3$, we first
calculate the power spectra, which is the absolute square of the
Fourier transform of the time sequence, for individual light curves. 
We assume the following fitting function for the power spectrum
$P(f)$, expecting the coexistence of the power-law component and the
thermal white-noise component.   
\begin{equation}
 P(f) = A f^{\alpha} + B .
\label{fit.eq}
\end{equation}

\begin{figure}[htb]
\epsscale{1}
\plotone{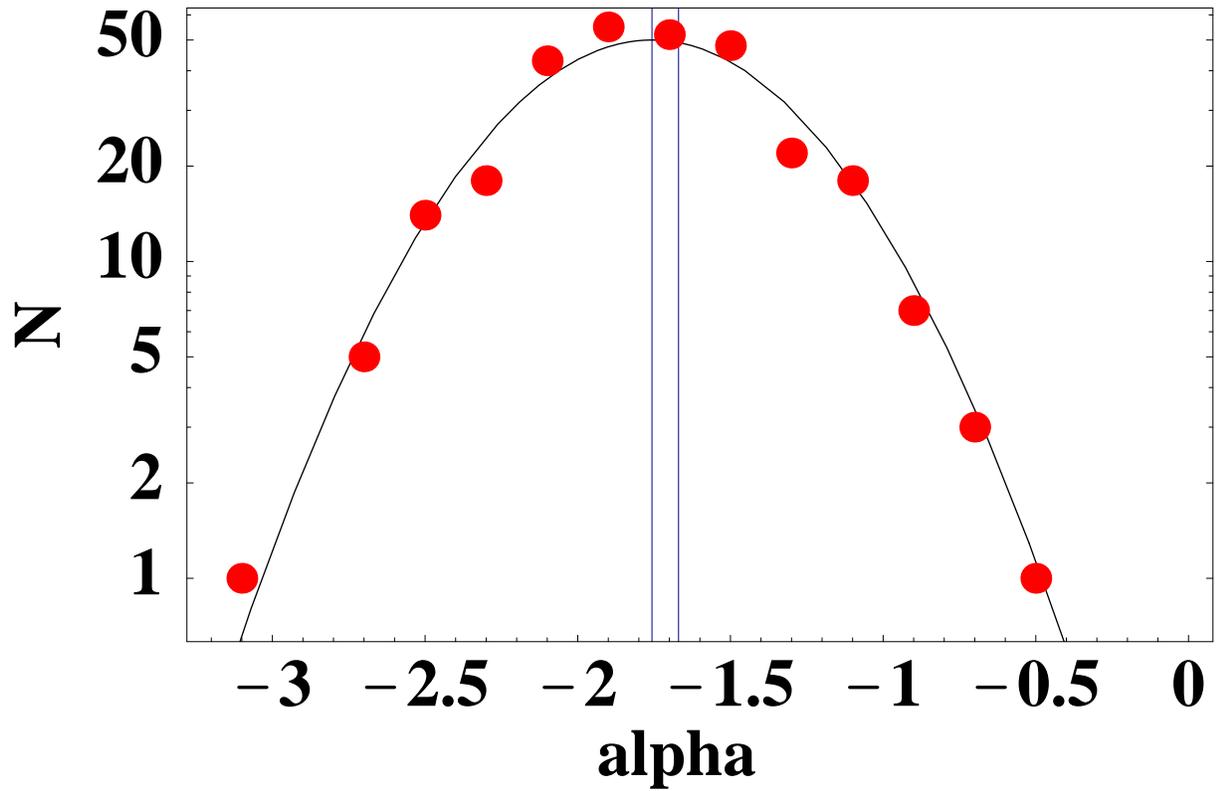}
\caption{The distribution of the power index $\alpha$ for 297
data sets. The distribution is well fitted by the Gaussian with the
mean $-1.76$ and the variance $0.65$.
}
\label{in.dist}
\end{figure}

Figure \ref{in.dist} shows the distribution of power-law index
$\alpha$ thus obtained.
The distribution is well fitted by the Gaussian and the mean
value is $-1.76$, which is close to the value $-1.67$ reported in
the reference \citet{Beloborodov98}.
We emphasize here that the individual data has the power-law PSD
though the power index is widely distributed with the variance 0.65.

\section{WHAT DETERMINES THE INDIVIDUAL POWER LAW? }

\label{origin}

Now let us consider the meaning of the power-law in PSD.
As is well known, the power-law itself in the PSD can be realized in
various origin.  The specifications of the origin is significant for
the analysis of the central engine of GRBs.  

\citet{Mineshige01} shows
at least the following three ways to realize
the power-law in PSD. 
\begin{enumerate}
\item  Specific global profile of a burst. 
\item  Levy-type random process.
\item  Superposition of self-similar shots.   
\end{enumerate}
We examine the above possibilities in the following in this order.  

Examining the first possibility, we realize that each GRB data generally
has multiple shots and rapid fluctuations as well as noise whose origin
is not seem to be GRB.
We try to identify these components in each GRB data step by step.  

First we discuss the light curve with a single shot. This simple
type of light curve shows a clear power-law behavior in its PSD.
We consider the following two model cases; 
a) the decay type, and
b) the grow-and-decay type.  

\noindent
a) The decay type

The decay type shot has the flux $x$ at time $t$ as  
\begin{equation}
x(t) = h t^{-p} \exp(-f_{0} t) \theta(t) ,
\label{fast.eq}
\end{equation}
where the positive parameters $h$, $p$ and $f_0$ respectively
represent ``intensity'',
``sharpness'' and ``inverse of the duration'' of the shot.  
$\theta(t)$ is the step function.  
Power spectrum $P$ for this shot is written as
\begin{equation}
P(f) = h^2[\Gamma(1-p)]^2 \left[ f_{0}^2 +f^2 \right] ^{p-1},
\end{equation}
where $\Gamma(x)$ is the Gamma function.

\noindent
b) The grow-and-decay type

The grow-and-decay type shot generally has asymmetric profile in
the burst; we separate the grow (left side) and
decay (right side) of the shot.
Each side of the shot is written as
\begin{eqnarray}
x(t)
 &=& h_L (-t)^{-p_L} \exp(t f_{0L}) \theta(-t)
   + h_R (t)^{-p_R} \exp(-t f_{0R}) \theta(t) 
\end{eqnarray}

Power spectrum of this type is written as
\begin{eqnarray}
P(f)
 &=& h_L^2 \{\Gamma(1-p_L)\}^2
 \left(f_{0L}^2 + f^2\right)^{p_L-1}
   + h_R^2 \{\Gamma(1-p_R)\}^2
 \left(f_{0R}^2 + f^2\right)^{p_R-1}
\nonumber \\
 & & {} + 2 h_L h_R \Gamma(1-p_L) \Gamma(1-p_R) 
\nonumber \\
 & & {} \times
 \left(f_{0L}^2 +f^2\right)^{\frac{p_L-1}{2}}
 \left(f_{0R}^2 +f^2\right)^{\frac{p_R-1}{2}}
 \cos \left[(p_L-1) \theta_L + (p_R-1) \theta_R\right] ,
\label{eq.gn_P}
\end{eqnarray}
where $\theta _{{\rm L,R}}^{}  = \arctan \left(f/f_{0{\rm L,R}}\right)
$.  

Since the GRB data generally has multiple shots, the above single
shot profiles must be superposed before we use. For simplicity
for fitting procedure, we assume that all shots in the individual
light curve has the same profile except the overall amplitude 
\footnote{
When we fit actual data of GRBs, we use inverse of duration $f_0$ 
as a fitting parameter, but it does not change the index of PSD at all.
}; i.e. the $k$-th shot in the light
curve $x_k (t)$ is written as $A_k x(t-a_k)$, where $A_k$ is the
relative amplitude and $a_k$ is the location of the $k$-th shot 
with $x(t)$ being the fixed function for each GRB data. 
Then we can easily calculate the PSD for multiple shots since the
Fourier transforms of the first and the k-th shots are simply
related with each other as $\tilde{x}_k = A_k e^{ifa_k} \tilde{x}_1$.  

The light curve $x_T(t)$ which has $n$ shots
\begin{equation}
x_T(t) = \sum_{k=1}^{n} x_k(t) ,
\end{equation}
has the PSD
\begin{eqnarray}
P_T(f)
&=& P(f) 
 \left[\sum_{j=1}^n A_j^2
 + \sum_{j>k} 2 A_jA_k \cos f(a_j - a_k) \right],
\end{eqnarray}
where $P(f)$ is PSD for $x(t)$.

It is apparent that the last expression consists of
the ``constant'' term
and the ``oscillating'' term; the former term determines the global
profile of PSD and the latter fluctuations around it. 

We can actually see this structure in PSD of GRB data. 
We first identify the maximum shot in the GRB data and locally fit
this shot by the single-shot form argued in the above.  Then we subtract
this first fit from the original data yielding one-shot
subtracted data. Second, we identify the maximum shot in this
subtracted data and locally fit this shot by the single-shot form
with the same parameter $p$.  Then we subtract this second
fit from the one-shot subtracted data yielding two-shots
subtracted data.  After repeating this process several
times, the reduced data becomes almost flat.  

\begin{figure}[htb]
\epsscale{1}
\plotone{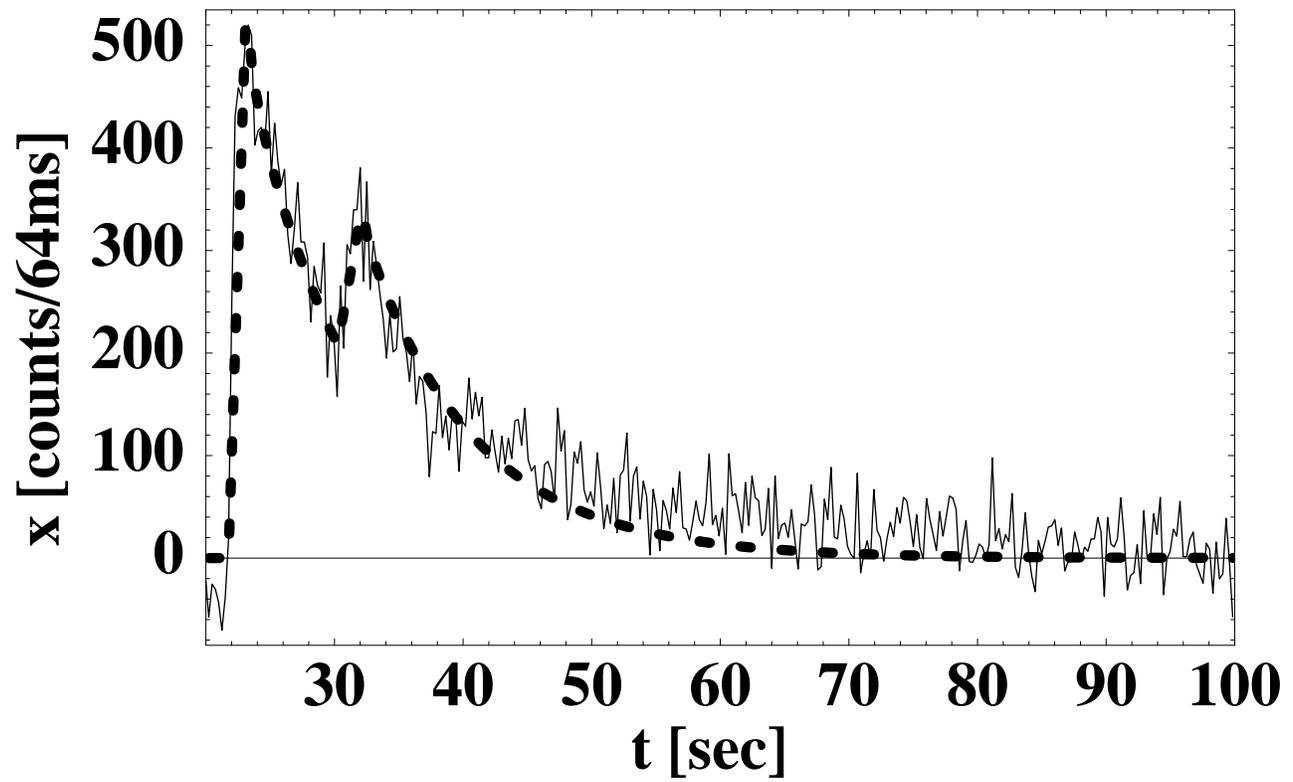}
\caption{
Actual light curve GRB910602(\#257) (solid line) and fitting
shots (broken line); the 
superposition of the first and the second fits.
The parameter $p$ is $0.03$.
}
\label{fit.lc}
\end{figure}

\begin{figure}[htb]
\epsscale{1}
\plotone{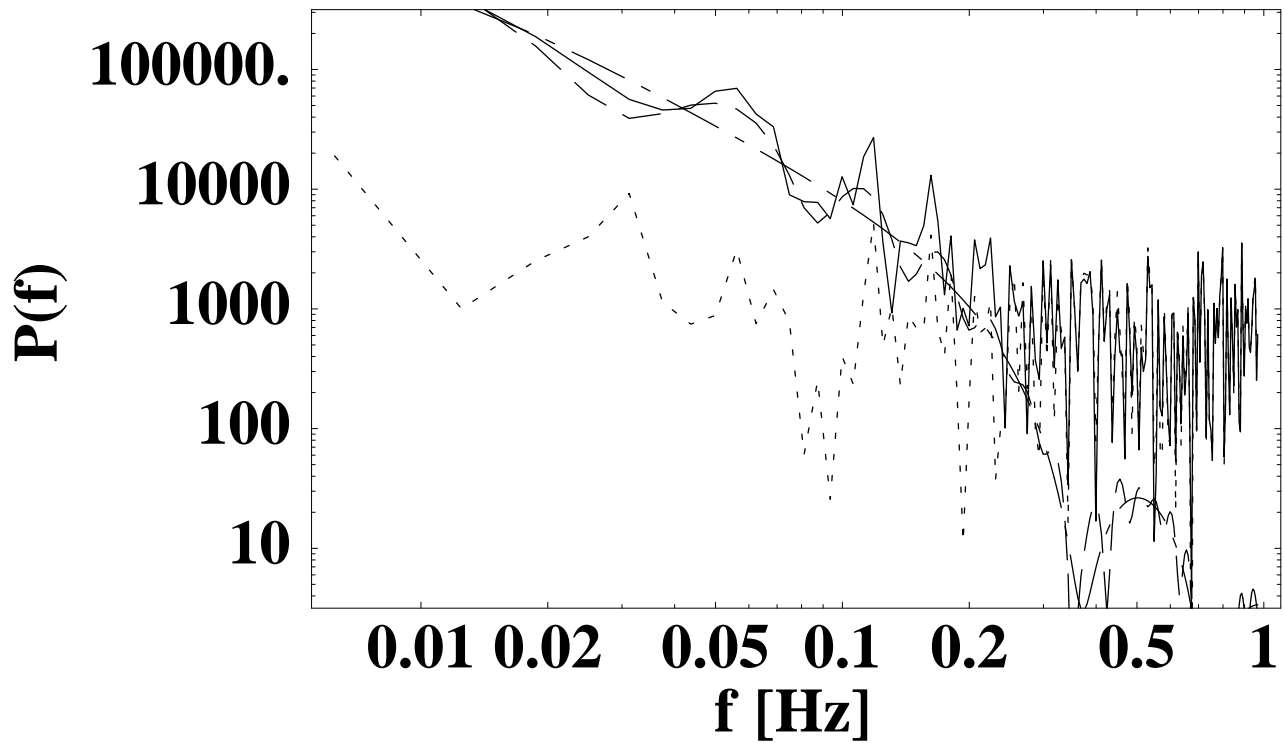}
\caption{
Power spectrum of the light curve in Fig.\ref{fit.lc}.
We plot PSD of actual data (solid line), PSD of the
first fit (chain line), PSD of the superposition of the
first and the second fits (broken line) 
and PSD of the two-shots subtracted data (dotted line).
PSD of the original data is hierarchically decomposed into the
first shot, second shot and 
the small fraction of the subtracted component; our method is
successfully working.  
Note that the PSD of the first fit already matches well
with that of the original data 
especially at low frequency regions.  
Power index of this data is $-1.95$.
}
\label{fit.pw}
\end{figure}

We applied this method for many individual GRB data.  
One typical example for GRB910602(\#257) is shown in Fig.\ref{fit.lc}.  
Here we use the decay type shot for fitting identifying two shots
in the data.  
The full PSD of this light curve is plotted in Fig.\ref{fit.pw} with the
solid line.  
The PSD of the first fit is plotted with the chain line which
already shows smooth power-law.  
The PSD of the superposition of the first and the second fits
is plotted with the broken line 
which shows the oscillation around the chain line.  
It is almost clear that the PSD of the superposition of first few
fits are sufficient to faithfully reproduce the main part
of the original PSD.  
Other GRB data sets show similar behavior as this one.  

Thus we have confirmed the first possibility: Specific global
profile of a burst determines the power-law behavior in PSD.

In order to exclude the second possibility, the Levy-type random
process, we have examined the peak distribution analysis.  
We calculate the distribution function of all peak-intensity
in the data of artificially produced 
Levy-type light curve and that in the actual data of GRBs.
The former shows a clear power law however the latter does not.  
We can now claim that the power-law in PSD of GRBs is not
originated from the Levy-type random process.  

In order to exclude the third possibility i.e. superposition of
self-similar shots, we calculate the PSD of the few-shots
subtracted data.  For the previous data set, we calculate the
PSD of the two-shots subtracted data in Fig.\ref{fit.pw} with
the dotted line.
This residual component is 10-100 times smaller than the main component
and is almost flat.
If GRBs have self-similar structure in their light curves, the few-shots
subtracted data should also show power law (but its range must
be narrower than that of original data).
Thus we can now claim that the power-law in PSD of GRBs is not
originated from the 
superposition of self-similar shots 

We have also applied the same analysis to the data set of the
soft gamma-ray repeaters (SGR). We use the data of
SGR1900$+$14
\footnote{
\url{http://ssl.berkeley.edu/ipn3/sgr1900$+$14.lightcurve}}.
The first one-shot fit has been enough to reproduce
the power-law behavior in its original PSD
\footnote{Fitting paramerer $p$ is $0.045$ and power index of original PSD is $-1.8$.}
and the one-shot subtracted data shows
a flat PSD except several very sharp peaks which are considered to
be originated from the central pulsar of SGR.  

\section{WHAT DETERMINES THE AVERAGED POWER LAW? }
\begin{figure}[htb]
\epsscale{1}
\plotone{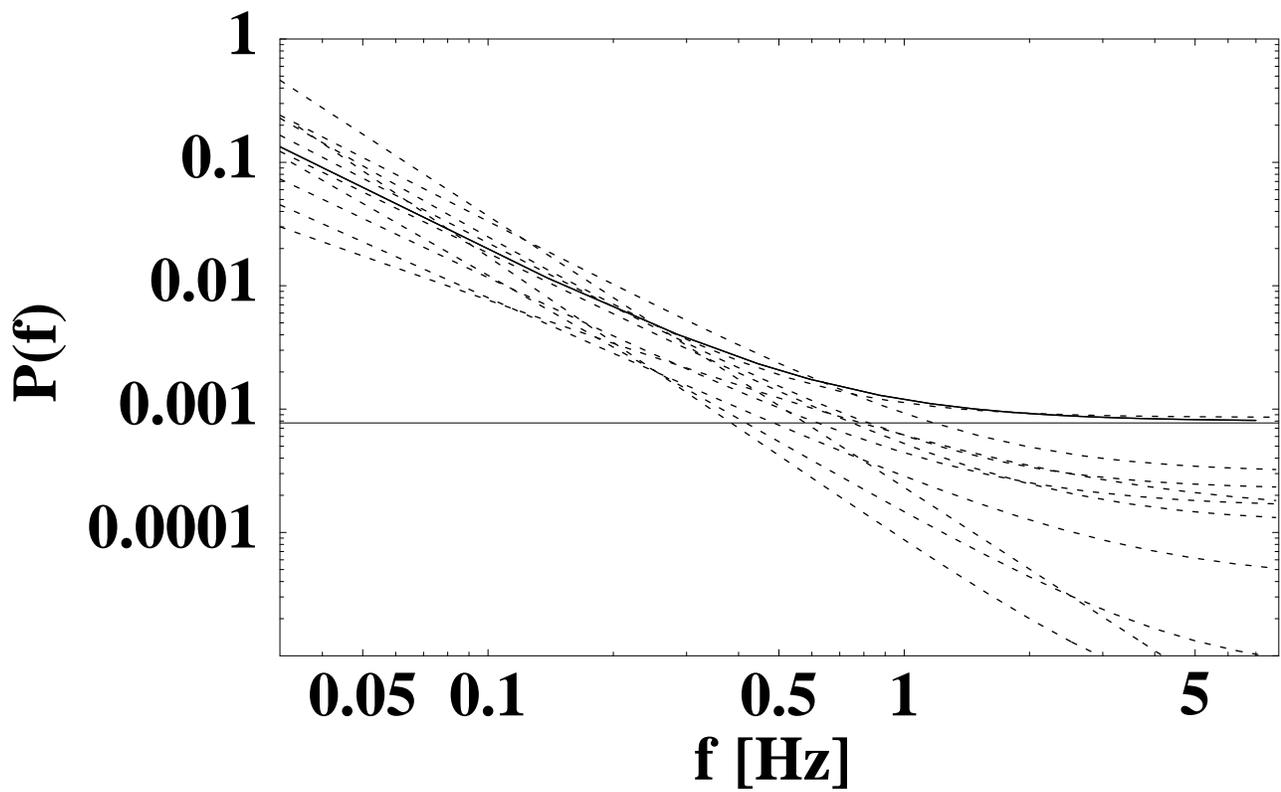}
\caption{Fits of individual (dotted lines) and average (solid line)
power spectra. We plot oldest 10 data of individual power spectra out of
297, and plot average of all 297 data.}
\label{fit.mu}
\end{figure}

We have averaged all the 297 PSD of GRB data and obtain
the Figure \ref{fit.mu}.   
Before taking the average, we have normalized each GRB luminosity data
so that the maximum of the count rate be unity. This is because the
original light curves have two to three orders of difference in maximum
count rates and therefore the naively averaged PSD is determined by
few GRB data. 
Since the magnitude of the count rate itself is not
physical (we don't know the distance!), such discrimination
is not a proper manipulation.   
We have also tried the total-count-rate (fluence) normalization as well
as maximum-count-rate normalization; yielding no significant
difference between them.  

Thus averaged PSD shows much clear power-law (Fig.\ref{fit.mu}) than
the individual PSD 
of GRB data.  $ $ From Fig.\ref{fit.mu} we again observe that the
individual PSD shows 
power-law and the power index simply fluctuates variously.  After
the superposition of 
many PSD data, there appears smooth power-law behavior with the
power index of the 
central value.  It is important to observe in Fig. \ref{fit.mu} that
there is no systematic 
correlation in each PSD to yield the global smooth power-law
behavior.  Especially there 
is no systematic distribution of time scales nor correlation of
turning points which, if any, 
would have yielded clear envelope when many PSD are superposed. 

The above reasoning for the appearance of the smooth power-law
in averaged
PSD is also supported by the fact that the power index $-1.67$ which
is close enough to 
the mean value $-1.76$ of the index distribution within 14 percent
of the variance.  
If the clear power-law were realized by the envelope of many
systematic distribution of 
burst time scales, the averaged power index would be significantly
smaller than the mean 
power index.

\section{CONCLUSIONS AND DISCUSSIONS}

Analyzing 297 power spectrum density (PSD) of GRBs, we obtain
the following results in this paper.
a) Individual GRB data shows power law behavior in the PSD.  The
power index ($\alpha$) distribution is well approximated by
the Gaussian form with the mean $-1.76$ and the 
variance $0.65$.  
b) Power law behavior in PSD for individual GRB data is determined
by the shot profile and 
not by the Levy-type noise nor the superposition of many
self-similar shots.    
c) Power law index in averaged PSD is simply determined by
the mean value of the index 
($\alpha$) distribution.  We found no correlation mechanism
for producing clear power law in averaged PSD.  

From the result b), we doubt the turbulence
\citep{Beloborodov98,Beloborodov00} as
the origin of the power law in PSD of GRBs.  This is because
the self-similar cascade of eddy, which would 
naturally yield many self-similar shots in the emission, yields
the Kolmogorov power law 
spectrum of velocity fluctuations in turbulent medium.    

The result b) provides sharp contrast with the case of power
law in the X-ray emission from AGN.  Many authors have demonstrated
that some self-similar cascade model, which 
yields many self-similar shots, successfully describes the
emission profile and the power law in PSD of AGN.  Thus we
naturally expect that the emission mechanism of AGN and 
that of GRB are quite different with each other despite the
similar power law behavior in their PSD.  

As we have seen in our analysis, some special form of
shots, which yield the power law in PSD, is very characteristic
and would be a good criterion for restricting various models of 
GRB generation mechanisms.  The special shot profile is more
severe checking point of the validity of the model than simply
the power-law behavior of PSD
\citep{psm99,spm00}.
An urgent interest then would be the question whether the
popular "Internal shock model" 
can explain these special shot profiles.  

There are still important questions to be answered in
the near future. 
1) The shots in the light curves of GRBs have distinctly
different two types; the decay type 
and the grow-and-decay type. This property is different from
that of blazars, even if their 
similarity in emission process is often suggested. 
2) What is the distribution of the sharpness parameter $p$ and the
ratio of power-law component and noise $A/B$ in
equation (\ref{fit.eq}) ? Is it also Gaussian 
or any other form from which we can extract the information
of the distance or classification of GRB?  

\acknowledgments

We all would like to thank Takashi Okamura for many useful suggestion.
I. J. would also like to thank Prof. T. Yokobori and
Prof. A. T. Yokobori, Jr for their hospitality.

\end{document}